\documentclass[manuscript]{acmart}

\AtBeginDocument{%
  \providecommand\BibTeX{{%
    \normalfont B\kern-0.5em{\scshape i\kern-0.25em b}\kern-0.8em\TeX}}}

\setcopyright{rightsretained}
\copyrightyear{2024}
\acmYear{2024}
\acmDOI{10.1145/1122445.1122456}

\newcounter{quotecounter}
\newcommand{\labeledquote}[1]{\refstepcounter{quotecounter}\label{#1}}
\newcommand{\qbox}[1]{\raisebox{0.3ex}{\colorbox{blue!20}{\scriptsize Q{#1}}}}

\usepackage{colortbl} % Color table rows, columns, and individual cells

\usepackage{todonotes}
\usepackage{makecell}

\usepackage{enumitem}
\usepackage{amsmath}
\usepackage{listings}
\usepackage{soul}

\begin{document}

\title{Supporting Design Education: Integrated Measurement and Presentation of AI-based Multiscale Design Analytics}
\title{Supporting Design Education: A Dashboard Indexing Multiscale Design Analytics to Design Instances to Demonstrate What is Measured}
\title{A Dashboard Indexing Multiscale Design Analytics to Design Instances: Supporting Design Education by Demonstrating What is Measured}
\title{Indexing Analytics to Instances: How Integrating a Dashboard can Support Design Education}

\author{Ajit Jain}
\authornote{Current affiliation: Audigent}
\email{ajain24@tamu.edu}
\author{Andruid Kerne}
\email{andruid@ecologylab.net}
\authornote{Current affiliation: University of Illinois Chicago}
\author{Nic Lupfer}
\authornote{Current affiliation: Mapware}
\author{Gabriel Britain}
\authornote{Current affiliation: Microsoft}
\author{Aaron Perrine}
\author{Yoonsuck Choe}
\author{John Keyser}
\author{Ruihong Huang}
\author{Jinsil Seo}
\affiliation{%
  \institution{Texas A\&M University}
  \country{USA}
}

\author{Annie Sungkajun}
\affiliation{%
  \institution{Illinois State University}
  \country{USA}
}

\author{Robert Lightfoot}
\author{Timothy McGuire}
\affiliation{%
  \institution{Texas A\&M University}
  \country{USA}
}

\renewcommand{\shortauthors}{Jain et al.}
\renewcommand{\shorttitle}{Indexing Analytics to Instances}

\begin{abstract}
We investigate how to use AI-based analytics to support design education. The analytics at hand measure \emph{multiscale design}, that is, students' use of space and scale to visually and conceptually organize their design work. With the goal of making the analytics intelligible to instructors, we developed a research artifact integrating a design analytics dashboard with design instances, and the design environment that students use to create them. 

We theorize about how Suchman's notion of mutual intelligibility requires contextualized investigation of AI in order to develop findings about how analytics work for people. We studied the research artifact in 5 situated course contexts, in 3 departments. A total of 236 students used the multiscale design environment.  The 9 instructors who taught those students experienced the analytics via the new research artifact.

We derive findings from a qualitative analysis of interviews with instructors regarding their experiences. Instructors reflected on how the analytics and their presentation in the dashboard have the potential to affect design education. We develop research implications addressing: (1) how indexing design analytics in the dashboard to actual design work instances helps design instructors reflect on what they mean and, more broadly, is a technique for how AI-based design analytics can support instructors' assessment and feedback experiences in situated course contexts; and (2) how multiscale design analytics, in particular, have the potential to support design education. By indexing, we mean linking which provides context, here connecting the numbers of the analytics with visually annotated design work instances.

\end{abstract}

%% Please copy and paste the code instead of the example below.https://www.overleaf.com/project/5ecd75b46a29c90001cd3612
%%



\maketitle

\section{Introduction}
We investigate supporting design education by directly linking AI-based analytics with instances of students' visual design work, in order to demonstrate what is measured.
Our approach integrates an analytics dashboard with a collaborative design environment, where students and instructors create and comment on design work instances.
The analytics at hand, multiscale design analytics, were introduced by Jain et al \cite{jain2021recognizing}.
\emph{Multiscale design analytics} measure how creative work, involving pictorial and textual elements, uses space and scale as organizing principles for presentation.

\begin{figure}[t!]
  \centering
  \includegraphics[width=\linewidth]{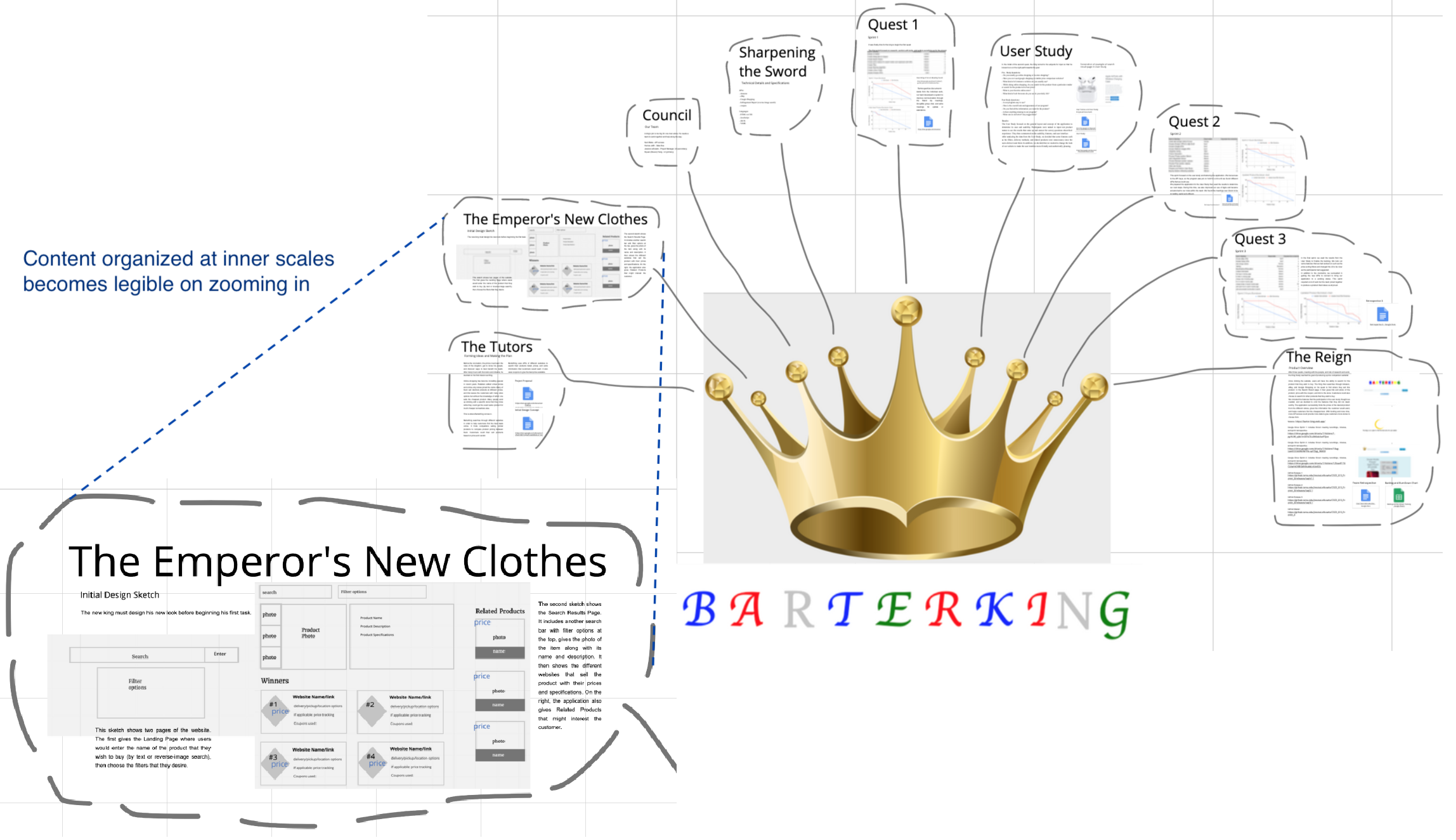}
  \caption{Multiscale design produced by a student in instructor I4's human-computer interaction project. The student organized different phases of the project as clusters of design elements, nested across scales of magnification. 
  At the outer scale, the crown, its caption, and spikes connected with each phase together form an overall cluster. 
  In the bottom left, we see the view as the user zooms in. Elements deeply nested in one scale become legible at the next.
  We observe nine nested clusters at the inner scale.}
  ~\label{fig:multiscale}
  \vspace{-2em}
\end{figure}

\textit{Multiscale design}, according to Lupfer et al., is ``the use of space and scale to explore and articulate relationships, [which] involves the juxtaposition and synthesis of diverse design elements'' \cite{lupfer2018multiscale} (See example in Figure \ref{fig:multiscale}.).
\textit{Scale} refers to levels of magnification:
elements at the same scale are ``equally legible at the same viewport zoom'' \cite{lupfer2019multiscale}.
In practice, we observe that the juxtaposition and synthesis of elements, across scales, tends to look like nested spatial groups, i.e., \textit{clusters} (Figure \ref{fig:multiscale}), of elements.
Multiscale design is a strategy for what Tufte calls \textit{escaping the flatland of envisioning information}, i.e., for increasing the legible dimensionality and density of information on the screen and page \cite{tufte1990envisioning}.
Prior investigations of design course contexts found that multiscale design supports students in creative ideation and reflection \cite{lupfer2019multiscale}, schematic development of design proposals \cite{lupfer2018multiscale}, and reuse of ideas across project deliverables \cite{britain2020}.

In the diverse design education contexts we study, students work on creative, open-ended projects, where there can be multiple approaches to solving a problem, i.e., there is no fixed right answer.
In fact, excellent answers are characterized by creativity, as well as skill, applied in the context of a particular design project problem.
Working on such projects is known to be challenging \cite{osmond2015threshold}.
Students need frequent assessment and feedback in order to make progress \cite{oh2013theoretical,krause2017critique}.
Confounding these needs, in practice, is the trend of growing design education demands. 
Larger class sizes confront instructors with challenges in providing timely assessment and feedback \cite{krause2017critique}. 
Prior work shows that the use of AI can complement human work by providing the ability to process big data at speed \cite{kittur2019scaling}.
The present research thus investigates AI-based multiscale design analytics as a means to support and scale instructors' teaching efforts.
Further, these analytics can form a basis for a constructivist learning model \cite{sin2015application,jonassen1994thinking}, where students learn, through making, even when instructors are not there.

Consider a scenario.
Alex is teaching a large class where students learn to use multiscale design.
As students work on projects and put multiscale design principles and skills into practice, it is vital to provide them with feedback in early stages.
However, due to time constraints, Alex is unable to continuously assess each student's creative design work in detail.
An AI-based system derives analytics that measure multiscale design characteristics in creative works, providing Alex with insights of how their students are using space and scale.
The analytics computed for each design work assist them in providing quick, personalized feedback to individual students.
Further, aggregate views of analytics\textemdash for each deliverable\textemdash assist Alex in identifying recurring problems students are facing in the course.
Based on that, Alex develops timely pedagogical interventions, which progressively help improve student work and overall course outcomes.

In meeting the needs suggested by this scenario, we draw on recent AI research, which has shown promising results in recognizing creative visual design characteristics, such as nested hierarchies in spatially organized content elements \cite{jain2021recognizing,oulasvirta2018aalto}.
The research gap we address is a lack of prior understanding of how measures based on AI recognition can support instructor assessment and feedback experiences in design education. 
Consideration of the demands of Suchman's precept of \emph{mutual intelligibility}, here, how an AI recognizer is understood by people who use it, suggests conducting such an investigation in situated design course contexts.
By combining understanding of multiscale design with AI recognition techniques, we are able to investigate how instructors respond to new forms of design education analytics.
Our research questions are:

\textit{RQ1: How, if at all, can AI-based design analytics support instructors' assessment and feedback experiences in situated course contexts?}

\textit{RQ2: What specific value can AI-based multiscale design analytics provide to design instructors in situated course contexts?}

AI-based design analytics are a relatively new technology, which is challenging to investigate, because performing studies in situated course contexts are demanding.
Ethics prescribe that it is necessary to provide all the functionalities sufficient to support instructors and students so we do not hinder learning \cite{brown1992design}.

In order to contribute new knowledge about how AI-based design analytics have the potential to be useful, we take a \textit{Research through Design} (RtD) \cite{gaver2012should,zimmerman2007research,zimmerman2010analysis} approach.
RtD is suitable when the goal is to explore ``what might be'' rather than a ``comparison or refutation'' of certain techniques.
As Zimmerman et al. explain, the focus of RtD is on making the ``right thing'', not making the thing right \cite{zimmerman2007research}.
Prior research \cite{rebanal2021xalgo} took a similar design probe approach to develop understandings of users' experiences with animations explaining an algorithm's functioning.
The present research takes an ethical RtD approach by developing a highly functional `research artifact' suitable for deployment in real world course contexts.

We studied the research artifact in 5 situated course contexts, in 3 departments, to understand how multiscale design analytics and their presentation can support instructors' assessment and feedback.
A total of 236 students used a multiscale design environment. 
The 9 instructors who taught those students experienced the analytics via the new research artifact dashboard integrated with the design environment.
We derive findings from a qualitative analysis of interviews with instructors regarding their experiences.
Based on the findings, we derive implications for making AI-based analytics useful in design and education. 
We conclude by reiterating the contributions of the present research and potential broad impacts.

\section{Prior Work}
We situate the present research amidst prior work involving learning activities, design, creativity, and learning analytics.
We begin by situating the student project tasks and courses within a broader framework of learning activities.
We follow with further consideration of how multiscale design matters in people's processes of organizing complex information.
Next, we consider approaches for assessing creative visual design, which is tricky, because there can be many approaches to solving a given design problem. 
Because our research involves computational assessment methods in design education, we then present prior work on the use of learning analytics and dashboards. 
%educational objectives of learning activities and 

\subsection{Educational Objectives of Learning Activities}
Bloom developed a taxonomy of educational objectives, organizing underlying cognitive processes in six categories: knowledge, comprehension, application, analysis, synthesis, and evaluation \cite{bloom1956taxonomy}.
The taxonomy is organized as a pyramid with cognitive processes under the `knowledge' category at the bottom and `evaluation' at the top.
As Armstrong describes, from bottom to top, objectives are organized in a continuum of simple to complex and concrete to abstract \cite{armstrong2016bloom}.
Krathwohl et al. revised the original taxonomy to distinguish the `knowledge' and `cognitive' dimensions of the objectives \cite{krathwohl2002revision}.
Knowledge dimension categories include factual, conceptual, procedural, and metacognitive.
Cognitive dimension categories include remember, understand, apply, analyze, evaluate, and create.

In the design course contexts we study,
students engage in multiscale design, as they work on open-ended projects involving iterative phases of requirement gathering, idea generation, prototyping, and evaluation.
These design phases engage students in learning activities that represent abstract and complex cognitive processes\textemdash at the top of the pyramid in both original and revised taxonomies\textemdash such as analyze, evaluate, and create.
Multiscale design\textemdash as a form of creative visual design\textemdash corresponds to the `create' category in the revised taxonomy, which is described as ``putting elements together to form a novel, coherent whole or make an original product.''

To effectively assess student learning, it is vital to develop measures aligned with educational objectives.
In engineering design, Summers and Shah advocate developing measures such as size, coupling, and solvability \cite{summers2010mechanical}.
According to them, these measures have the potential to assist instructors in assessing students' ability to address complex design problems
In the design course contexts we study, multiscale design is significant.
Hence, we develop computational measures for how students use space and scale in their design work.
These measures have the potential to assist instructors in assessing students' ability to put diverse elements together into a new, coherent whole.

\subsection{Multiscale Design}
Ray and Charles Eames demonstrated how we humans conceptualize our knowledge of the universe across powers of ten \cite{eames1968powers}.
Tufte articulates micro / macro readings, a strategy for constructing data narratives in which designers employ ``fine details'' that accumulate to form ``larger coherent structures'' \cite{tufte1990envisioning}.
Perlin and Fox conceptualized and actualized the importance of organizing information across scales, in the form of a zoomable user interface (ZUI) \cite{perlin1993pad}.
Bederson discusses how a ZUI helps people in developing a mental map of the information by taking advantage of human memory and spatial perception \cite{bederson2011promise}.

Design environments supporting multiscale design\textemdash e.g., Photoshop, Illustrator, InDesign, and IdeaM{\^a}ch{\'e} \cite{lupfer2019multiscale,lupfer2016patterns}\textemdash enable going beyond 2D by organizing content across scales, i.e., across a range of zoom levels.
Bar-Yam describes multiscale design as an approach to manage increasing complexity \cite{bar2006engineering}.
According to Barba, multiscale design supports a holistic analysis and development of ideas, across multiple scales, by allowing people to shift cognitive point of view up and down hierarchies \cite{barba2019cognitive}.
As Alexander describes, by employing hierarchies, designers organize a component both as a unit and as a pattern consisting of other units, and thus address the important challenge of building up a form from components \cite{alexander1964notes}.

Lupfer et al.'s investigation of a landscape architecture classroom found that multiscale design pervades student projects \cite{lupfer2018multiscale}. 
In these projects, multiscale design supports students in schematically forming design proposals to meet the situated needs of sites involving waterways and land use.
In a study involving computer science courses, multiscale design has been found to support students' iterative and reflective ideation \cite{lupfer2019multiscale}.
In a study involving interactive art and design course contexts, multiscale design has been found to enable students in referring to and reusing ideas across project deliverables, which facilitates consistency in their work \cite{britain2020}.
The present research investigates AI-based multiscale design analytics to support instructors in assessing student work.
This is part of a broader problem of how to assess design work \cite{jain2021support}.

\subsection{Assessing Creative Visual Design}
As Gero and Maher discuss, in creative design, differences in the representation of ideas are a rule rather than an exception \cite{gero1993modeling}.
This makes assessing creative design challenging.
A common pedagogical tool is to organize ``design critiques'', where instructors, peers, and invited jury provide students with feedback on their work \cite{oh2013theoretical,dannels2011students}.
Feedback is based on a broad range of criteria, across dimensions, such as product, process, content knowledge, and communication \cite{de2009assessment}.
Criteria for product visual design focus on characteristics such as color, form, composition, and layout \cite{de2009assessment}.
However, instructor-led assessment is not able to keep up with the growing demands of design education \cite{krause2017critique}.

Approaches developed by creativity and crowd researchers align with visual design assessments in courses.
Kerne et al. developed creativity metrics for assessing information-based ideation tasks and activities \cite{kerne2014using}.
Their `Visual Presentation' metric includes criteria such as whitespace, alignment, and organization of ideas in lines, grids, or other shapes.
Human raters applied these metrics to assess free-form creative assemblage of ideas. 
Xu et al. developed guidelines for crowd assessment of visual design work, including criteria of proximity, alignment, repetition, and contrast \cite{xu2015classroom}. 
For multiscale design, Lupfer et al. measured the number of scales used, by counting the number of times one needs to zoom in, in order to make inner elements legible \cite{lupfer2019multiscale}.
Despite the potential of these approaches to assist design courses, they face similar limitations as instructor assessment, i.e., human support may not always be available.

Computational approaches have the advantage of processing data at speed and providing on demand assessment \cite{jain2017measuring}.
Reinecke et al. assessed website aesthetics by developing a regression model based on attributes such as color, symmetry, and the number of images and text groups \cite{reinecke2013predicting}.
Oulasvirta et al.'s Aalto Interface Metrics web service is aimed at providing assessments of a graphical user interface design, to help designers in identifying and addressing the shortcomings \cite{oulasvirta2018aalto}.
For multiscale design, Jain et al. developed a computational model based on spatial clustering, which identifies scales and clusters present in design work \cite{jain2021recognizing}.
However, prior computational approaches did not focus investigation on experiences in design education course contexts.

Our investigation presents data about instructor experiences with AI-based analytics by invoking Jain et al.'s model \cite{jain2021recognizing} to compute multiscale design analytics.
We consider prior work on the derivation and presentation of learning analytics in order to subsequently inform the approach we take for presenting multiscale design analytics in our research artifact.

\subsection{Learning Analytics and Dashboards}
Learning analytics and dashboards technologies have been found to support instructors in identifying student problems and intervening, which improved student retention and success \cite{arnold2012course}.
In lecture-based course contexts, analytics\textemdash e.g., the number of times a student accessed a resource, time spent, and length of textual annotations\textemdash have assisted instructors in assessing student understanding \cite{dawson2012using}.
Likewise, dashboards have proven effective in lecture-based contexts, providing a quick understanding of student progress through representations such as tables and graphs \cite{duval2011attention,verbert2014learning}. 

Design course contexts involve project-based learning \cite{Dym2005}. 
As Blikstein discusses, in project-based contexts, there is a need to measure more open-ended and complex characteristics, which can provide instructors with insights into students' creative processes \cite{blikstein2011using}.
In design course contexts, specifically, Britain et al.'s study surfaced this need, as they presented Fluency analytics\textemdash i.e., the number of elements, words, and images\textemdash to design instructors \cite{britain2020}.
While the instructors found Fluency useful in gaining insight into students' efforts across various dimensions, they desired more sophisticated analytics.
In our investigation, we present AI-based analytics to instructors\textemdash the number of scales and clusters\textemdash providing them insights into students' multiscale design organization.

Prior analytics dashboards focus on presenting facts.
But with advances in AI and its applicability in diverse domains, the community has begun researching AI-based analytics, which include inferences \cite{how2020artificial,lim2018design}.
Presenting AI results in a comprehensible manner is vital, so that users can trust the system and rely on its assistance \cite{shneiderman2020human}.
Among prior work, Oulasvirta et al. provide visualizations of assessed visual design characteristics\textemdash e.g.,  of clutter, colorfulness, and white space\textemdash within website design. 
However, they did not assess or present multiscale design characteristics.
The present research focuses on conveying the meaning of scales and clusters analytics computed with AI recognition.
For this, we integrate the dashboard presentation of analytics with the actual design work they measure.

\section{Research Artifact / Probe}
To enable the present research, we developed a
research artifact\textemdash in the form of a functional prototype\textemdash extending the \textit{technology probe} \cite{hutchinson2003technology} methodology.
We developed this research artifact probe in order to: (a) understand instructors' needs and desires in a real-world setting, (b) collect data through field-testing of AI-based analytics and their presentation, and (c) stimulate instructors' and our own thinking about technological possibilities. 
We did not strictly follow the technology probe methodology, in that we did continuously revise our prototype in an iterative design process and in that the underlying multiscale design environment possesses more extensive functionality than typifies technology probes.
The functionality we added for the present study, supporting multiscale design analytics and their presentation, was focused and limited in the spirit of technology probes.

We present a research artifact probe presenting AI-based multiscale design analytics with an analytics dashboard linked to design instances that can be viewed and edited in a free-form, multiscale design environment.
Through this integration, we sought to make relationships visible between the analytics and particular design element assemblages that they measure.

\begin{figure}[!t]
  \centering
  \includegraphics[width=0.7\linewidth]{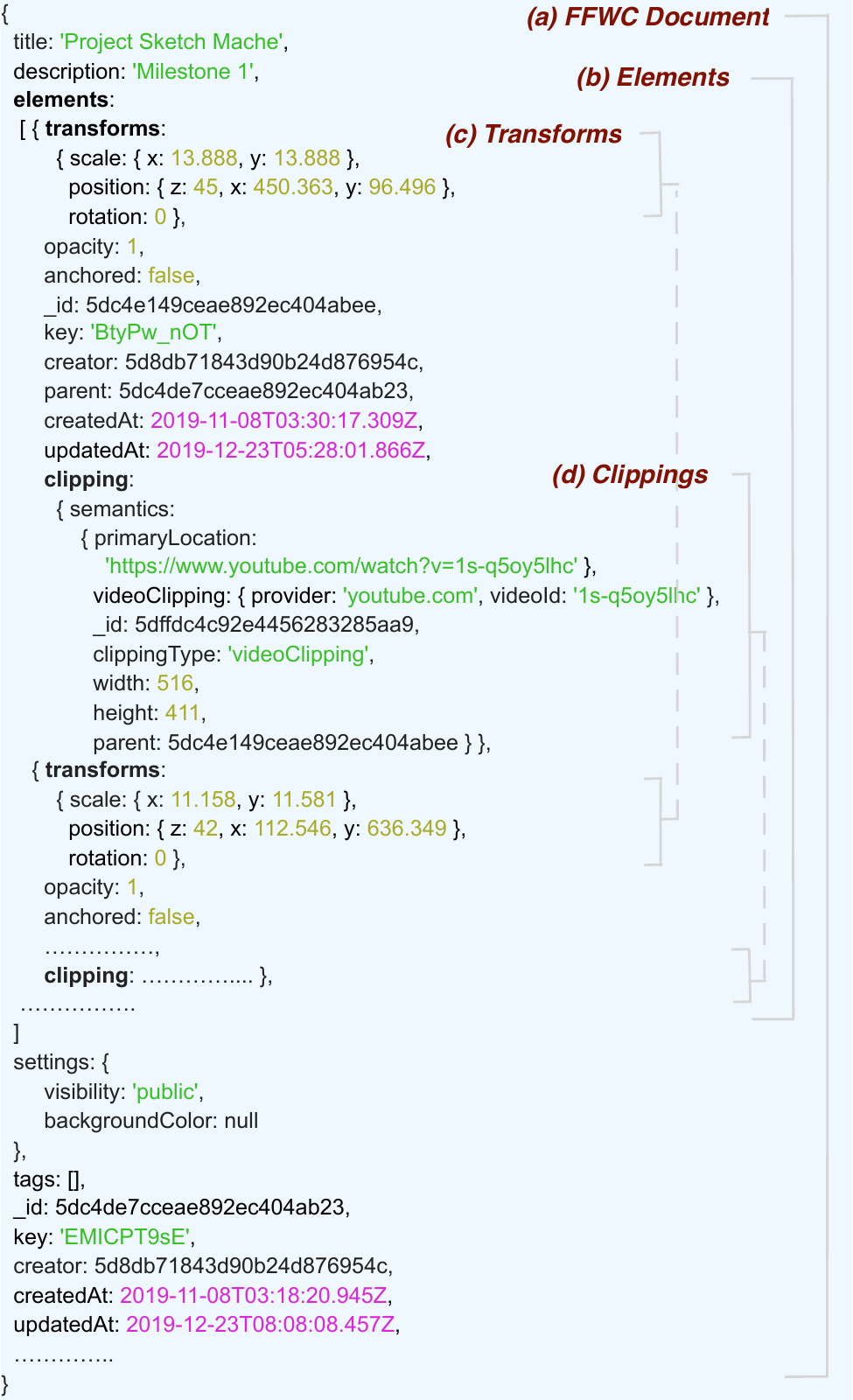}
  \caption{Each multiscale design document\textemdash a free-form web curation (FFWC)\textemdash is stored as JSON \cite{jain2021recognizing}. It is comprised of properties such as \textit{title}, \textit{description}, \textit{creator}, and a collection of \textit{elements}. Each element includes a \textit{transforms} property\textemdash with sub-properties position, scale, and rotation\textemdash which allows determining the element's spatial location with respect to the origin. Each element also stores a \textit{clipping} property, which in turn stores semantics extracted at the time of collecting the element.}
  ~\label{fig:ffwc_document}
  \vspace{-2em}
\end{figure}

\subsection{Multiscale Design Environment}
\label{sec:multiscale_environment}
LiveM{\^a}ch{\'e} is free-form, multiscale design environment, which runs in a web browser and integrates an unusual set of capabilities \cite{hamilton2018collaborative}. 
It is intended to support real-time collaborative  ideation, by helping users discover and interpret relationships through visual and conceptual thinking and composing a whole.
The environment enables users to act as designers, collecting elements through direct drag and drop from web pages and creating elements through text editing and sketching.
Heterogeneous media elements can be directly collected, including images, text chunks, YouTube and Vimeo videos, Google Docs, Google Sheets, and Google Maps.
Assemblage of elements is supported, in a manner similar to Adobe Illustrator, through positioning, resizing, and layering.

More specifically, the multiscale design environment supports designers in invoking the following creative strategies: collect, assemble, sketch, write, shift perspective (pan and zoom), and exhibit \cite{kerne2017strategies}.
Designers \textit{collect} elements by dragging and dropping a variety of content from the web, e.g, from news, e-commerce, and social media web pages.
The environment extracts and stores semantic information associated with each collected element \cite{qu2014metadata}.
The semantic information at least includes the title and URL of the page the element is collected from.
Depending on the source, semantics may include additional information. 
For example, for a news article, the additional information includes authors and publication date.

Designers \textit{assemble} the collected and self-made elements through transformation operations, such as position, resize, rotate, layer, crop, and blend.
These transformations support designers in developing the representation of the whole.
Designers invoke the \textit{shift perspective} strategy, through pan and zoom, in order to navigate across space and scale.
Zooming out allows accessing ``encompassing contexts'' and zooming in allows navigating to ``nested details'' \cite{lupfer2018multiscale}.
Designers annotate ideas through \textit{sketching} and \textit{writing}.
Each multiscale design is assigned a unique URL, through which designers \textit{exhibit} their work.
Figure \ref{fig:multiscale} shows an example work produced using the environment.

The environment stores design representations in the cloud, as JSON, including semantic information, such as position (x and y), width, height, and any transforms applied to the constituent elements (Figure \ref{fig:ffwc_document}).
As Jain et al. \cite{jain2021recognizing} describe:
[The multiscale design document] consists of a collection (Figure \ref{fig:ffwc_document}.a) of content elements (Figure \ref{fig:ffwc_document}.b), each with graphical transformations (Figure \ref{fig:ffwc_document}.c). 
The document level (Figure \ref{fig:ffwc_document}.a) also stores properties, including title, description, key (used in web URL), id (a unique internal identifier), settings (visibility and background color), and creator. [The design environment] associates the extracted semantics with a reference to the collected element, which is together referred to as the \textit{clipping} within an element (Figure \ref{fig:ffwc_document}.d).

\begin{figure}[!t]
  \centering
  \includegraphics[width=0.85\linewidth]{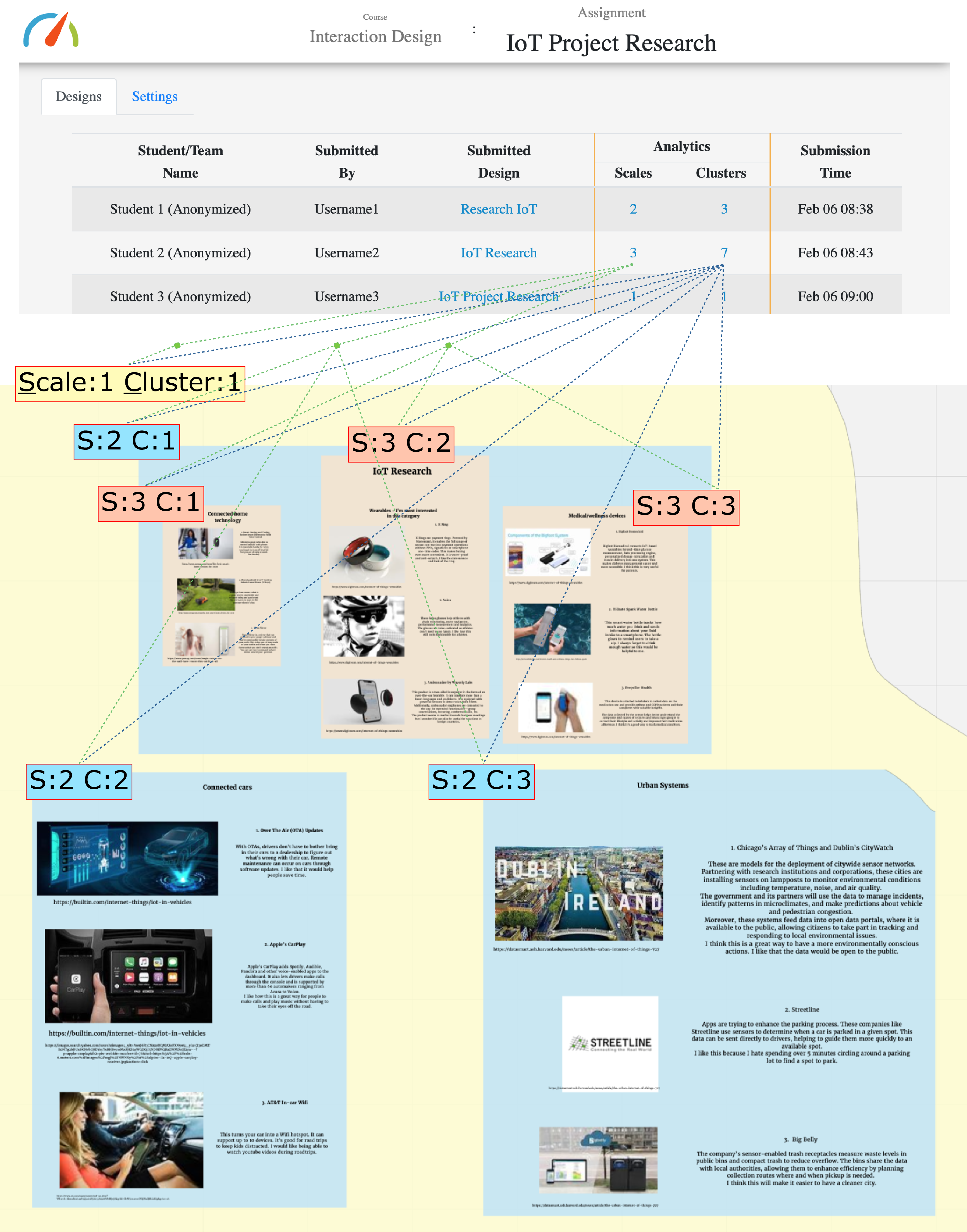}
  \caption{We integrated a dashboard with the underlying design environment with the goal of conveying the meaning of analytics to users. Clicking an analytic presented on the dashboard interface opens the actual design environment and shows the corresponding AI identified nested scales, with all clusters at a particular scale rendered in the same background color. In the above figure, the outermost scale comprises one cluster\textemdash including all design elements\textemdash which is rendered in yellow color. The next inner scale has three clusters (one at top and two at bottom), which are rendered in blue. The innermost scale has three clusters\textemdash within the top blue cluster\textemdash which are rendered in brown. The visualization makes relationships visible, between particular design element assemblages and analytics that describe and measure them. It enables instructors to understand what the analytics mean. }
  ~\label{fig:visualization}
  \vspace{-2em}
\end{figure}

\begin{figure*}[t!]
  \centering
  \includegraphics[width=\linewidth]{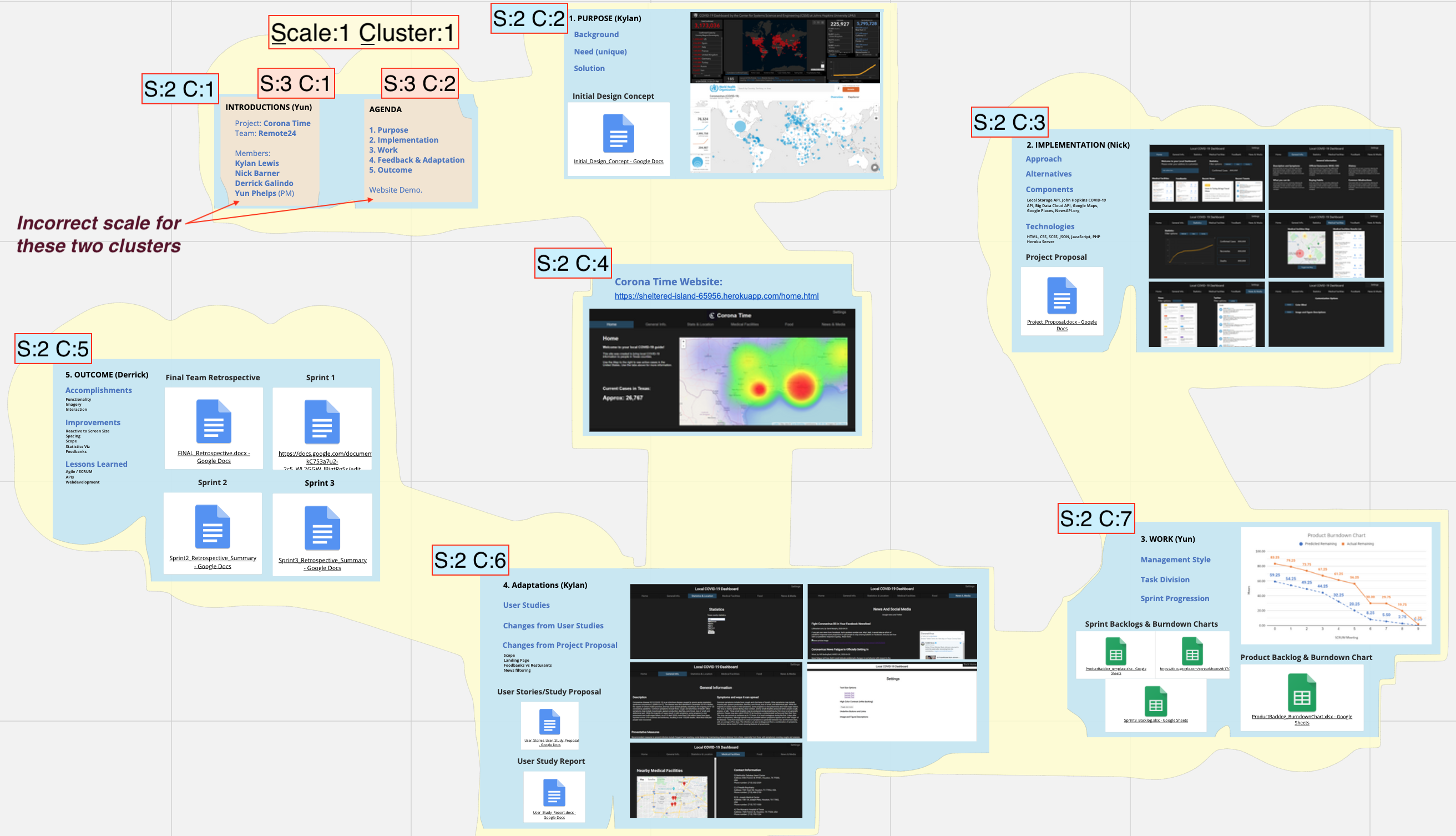}
  \caption{Another example of scale and cluster recognition. The outermost scale comprises one cluster\textemdash including all design elements\textemdash which is rendered in yellow color. The next inner scale has seven clusters, which are rendered in blue. However, AI is not perfect. Here, the two clusters rendered in brown are incorrectly recognized to be at the next inner level, as their content is similarly legible as the content within the clusters rendered in blue.}
  ~\label{fig:visualization2}
  \vspace{-2em}
\end{figure*}

\subsection{Integrating a Design Analytics Dashboard with the Multiscale Design Environment}
In order to connect design assessment with design work, we integrated multiscale design analytics into the multiscale design environment.
To contextualize the analytics, we link them with design work instances.
The multiscale design analytics that we present to instructors are (1) the number of scales and (2) the number of clusters present within a given design work instance.
Our linkage of design analytics to design instances makes visible to users what is being measured by analytics.

We choose an off-the-shelf AI model suitable for computing multiscale design analytics.
As we described in the introduction, our goal is to explore the potential of AI-based multiscale design analytics for supporting design instructors, not to compare or refute particular techniques.
We invoke Jain et al.'s \cite{jain2021recognizing} spatial clustering based model, as it has been previously used in design education contexts, for recognizing clusters of elements nested across scales within multiscale design work.
We input position, width, height, and transforms of multiscale design elements\textemdash extracting them from the JSON representation\textemdash and obtain scales and clusters as output from the model.

We integrate a design analytics dashboard directly with the multiscale design environment in which students perform their work.
As instructors regularly use the environment to 
view and assess student projects, the dashboard integration streamlines the process of making AI-based analytics available in their situated work contexts.
The dashboard interface allows instructors to manage a course, its assignments, and submission to the assignments.
When an instructor views an assignment page (see Figure \ref{fig:visualization} top) on the dashboard, it presents to them the students' submissions and scale and cluster analytics derived through the AI analysis of submissions.

We link the dashboard presentation of the analytics with the specific design assemblages they measure.
This makes results of AI recognition\textemdash for each instance of student design work\textemdash visible to the instructor (see Figure \ref{fig:visualization} bottom).
When the instructor clicks a scale or cluster analytic, it opens the corresponding student design work instance within the multiscale design environment.
An animation shows them the design assemblages for each scale and cluster.  
All clusters at the same scale are presented using the same background color, so as to help discern nested relationships (Figure \ref{fig:visualization} bottom, Figure \ref{fig:visualization2}).

\section{Methodology and Context}
\label{sec:methodology_and_context}
We present a methodology of deploying the research artifact / probe in situated course contexts, gathering qualitative data through instructor interviews, and performing qualitative analysis with a methodology loosely based on Charmaz's approach \cite{charmaz2014constructing} to grounded theory. 
Grounded theory refers to a family of qualitative ``constant comparative'' research methods, which involve collecting data, through techniques such as observation and interviews, transcribing the data, unitizing the data, grouping the units based on common interpretations of what they mean, naming the groups (aka `coding'), and developing theory from  codes and categories that emerge through this iterative process \cite{birks2015grounded}.

There are various methodological and philosophical approaches to grounded theory.
As Birks and Mills articulate, among Charmaz's contributions to the pool of grounded theory methodologies is ``a focus on the place of the author in the text, their relationship with participants, and the importance of writing...'' \cite{birks2015grounded}. 
We disclose our positionalities in this investigation.
The technology developers also played a principal role in the gathering and analysis of data.
Further, the developers worked not as detached scientific observers of the courses;
rather, we worked in collaboration with instructors. 
Our collaboration involves the co-creation \cite{sanders2008co} of pedagogy, technological capability specifications, and interaction design.
Both developers / qualitative researchers and instructors are among the authors of this paper. 

{
\renewcommand{\arraystretch}{1.3} % vertical padding
\begin{table*}[t!]
\center
\footnotesize
\begin{tabular}{c|c|c|c|c|c}
{\bf ID} & {\bf Gender} & {\bf Course} & {\bf Field} & {\bf Role} & {\bf Semester}\\
\hline
I1  & F & \makecell{Digital Media Design, \\ UI/UX for Games} & Interactive Art \& Design & Course Instructor & Spring 20 \\ \rowcolor[gray]{0.95}
I2  & F & Interaction Design & Interactive Art \& Design & Course Instructor & Spring 20 \\
I3  & F & Engineering Design & Mechanical Engineering & Course Instructor & Summer 20 \\ \rowcolor[gray]{0.95}
I4  & M & Programming Studio & Computer Science and Engineering & Course Instructor & Spring 20 \\ 
I5  & M & Programming Studio & Computer Science and Engineering & Course Instructor & Spring 20 \\ \rowcolor[gray]{0.95}
I6  & F & Programming Studio & Computer Science and Engineering & Lab Instructor / TA & Spring 20 \\
I7  & M & Programming Studio & Computer Science and Engineering & Lab Instructor / TA & Spring 20 \\ \rowcolor[gray]{0.95}
I8  & M & Programming Studio & Computer Science and Engineering & Lab Instructor / TA & Spring 20 \\
I9  & M & Programming Studio & Computer Science and Engineering & Lab Instructor / TA & Spring 20 \\
\end{tabular}
\caption{Five design course professors and four teaching assistants in the role of lab instructor interacted with AI-based analytics, presented via the dashboard integrated with multiscale, free-form design environment.}
\label{tab:participants}
\vspace{-3em}
\end{table*}
}

\subsection{Course Contexts}
We investigated the research artifact / probe (Section~\ref{sec:multiscale_environment}) in 5 course instances, across three departments, during Spring and Summer 2020 (Table \ref{tab:participants}).
These design course contexts are diverse.
We contextualize the design tasks that students perform in these courses.
In order to depict the creative work that students in this study are performing, which involves multiscale design, we first present overviews of each project. 
To detail the design tasks student perform through project phases, we then elaborate on I2's project deliverables.

Instructors' assessment plays a vital role, as students work through the project phases.
Frequent assessment helps students in making continuous progress \cite{oh2013theoretical} and fulfilling project and overall course objectives.
As project ideas and deliverables are organized within the multiscale design environment, it becomes a one-stop place for instructors to regularly monitor and assess student work.
The integration of dashboard with the environment, thus, weaves multiscale design analytics into instructors' situated contexts, in concert with their teaching and assessment processes.

\subsubsection{Assignment Overviews Across Fields}
Here are overviews of multi-week assignment sequences in 5 course contexts, across fields (Table \ref{tab:participants}).
The first two assignments are in Interactive Art and Design.
The next is in Mechanical Engineering.
The last two are in Computer Science and Engineering.

I1 (Interactive Art and Design): The students' task is a team-based, 4-week game interface design project.
Students research an assigned game genre and then sketch their own UI while keeping aesthetics in mind. 
The instructor introduces [the multiscale design environment (Section \ref{sec:multiscale_environment})] in the course, through which students organize and present at least 5 game examples, explanations for each on their UI design, mechanics, and any similarities.
Alongside their sketches, students need to use text annotations, to explain their ideas and thought processes.

I2 (Interactive Art and Design): The students' task is a team-based, 6-week interactive installation design project to improve people's experiences of artworks.
Students research, conceptualize, develop, evaluate, and present an interactive projection mapping based on physical computing techniques. 
The instructor introduces [the multiscale design environment] in the course, through which students organize and present inspirational ideas, the project description and plan, concept sketches, storyboards, and visuals of circuitry and interactive functionality.

I3 (Mechanical Engineering): The students' task is a team-based, 2-week analogy formation project.
Students identify a topic and then generate solutions by utilizing analogies across a semantic word tree. 
The instructor introduces [the multiscale design environment] in the course, through which students organize and present their topic, web searches, identified ideas, and analogy formation using the word tree method.

I4 and I5 (Computer Science and Engineering): The students' task is a team-based, 6-week web application project.
Students conceptualize, develop, evaluate, and present a mashup website, which needs to utilize at least three diverse web services. 
The instructors introduce [the multiscale design environment] in the course, through which students organize and present the project description, interface sketches, backlogs, burndown charts, findings from the user study, and videos of the functional product.

\subsubsection{I2's Interactive Art and Design project deliverables}
Here we elaborate on one of the projects.
This project consists of 4 deliverables, due over 6 weeks, where students develop an interactive installation design project.
\begin{enumerate}
    \item 
Students perform technical and concept research, create a detailed project description and plan, and develop concept sketches and storyboards, while focusing on gestures, usability, and user experience.
In [the multiscale design environment], they need to include at least three inspirational ideas, two interactive artworks, and types of inputs, outputs, and circuitry they plan to use.  
    \item 
Students prepare their circuit connection, with an Android phone, using ProtoPie (a physical computing toolkit).
In [the multiscale design environment], they need to include visuals that demonstrate a working connection.
    \item 
Students develop a lo-fidelity prototype for ProtoPie design, which should include color and typography.
In [the multiscale design environment], they need to include visuals of the design and functionality.
    \item 
Students work on developing a hi-fidelity prototype, conducting a user study, and presenting their projects. 
In [the multiscale design environment], they include visuals showcasing users' experiences with the prototype and the final presentation document.
\end{enumerate}

\subsection{Instructor interviews}
The research artifact / probe was used by 5 design course instructors and four teaching assistants (TAs).
We conducted semi-structured interviews with the professors and teaching assistants regarding their experiences at the end of the respective courses.
We asked instructors regarding: whether the analytics dashboard showed them anything new about students' learning;
whether and how they utilized analytics for monitoring, intervening, and assessment and feedback;
their thoughts on making analytics available to students on demand; and
their understanding of analytics and how seeing analytics' relationships with the actual design work affected their experiences. 
The complete set of interview questions can be seen in Appendix \ref{sec:questions}.

Invoking Charmaz's method for grounded theory qualitative data analysis \cite{charmaz2014constructing}, two authors first performed initial coding of three interview transcripts. 
They met to bring their initial codes into alignment, and formed tentative categories.
Then, they performed focused coding of the remaining interview transcripts, revising codes and categories, as needed, to suitably represent the salient phenomena.
We present the categories, including participant quotes illustrating the phenomena, in the next section.

\section{Findings}
\label{sec:interviews_qual_analysis}
We present findings from our grounded theory qualitative analysis of instructor interviews, regarding their experiences with the research artifact / probe.
We developed four categories through the analysis of interview data, which illustrate
(1) analytics providing insights and informing pedagogical action,
(2) support for exploration, understanding, and validation of analytics,
(3) use of analytics for assessment and feedback, and
(4) analytics supporting students' self-reflection.

\subsection{Gaining Insights and Informing Pedagogical Action}

Instructors in our study reported that multiscale design analytics provide them with novel and useful insights.
I6 compared the experience with learning management systems, such as Canvas and Blackboard.
According to I6, the scale and cluster analytics offer unique insights, which they have not encountered on any other system.
According to I1 and I5, the scale and cluster analytics help them understand students' progress on their design projects. 
I2 finds the analytics particularly useful in understanding how students have developed and presented structure in their design.

I1: \textit{I think using the dashboard and using the analytics is really helpful for me to kind of get an understanding of what [students are] doing.}

I2: \textit{I've been thinking like, you know, [scales and clusters] could be a very useful information for me, you know in terms of how students develop structure and present that structure at different levels.}

We find initial evidence for the value of insights\textemdash provided by multiscale design analytics\textemdash as a basis for pedagogical intervention \cite{lockyer2013informing,shum2012learning}.
I9 expressed that these analytics can help them find out whether students are able to effectively use the multiscale design environment, and take action, adjusting the curriculum, as needed.

I9: \textit{If there are multiple scales and clusters...they are at least using the environment efficiently. So if this number is extremely low for everybody...then maybe you need to [give] a tutorial on [the design environment]}.

\subsection{Support for Exploration, Understanding, and Validation of Analytics}

Aiding comprehensibility is an important challenge to address in developing user interfaces for AI-based technologies \cite{shneiderman2020human}.
AI-based multiscale design analytics represent complex characteristics of students' design work.
Instructors found that our dashboard design helps them explore and understand multiscale design analytics.
In particular, instructors expressed that the links on the dashboard, in conjunction with visual annotations about how the algorithm operated on design instances, help them explore and understand the relationships between analytics and the scales and clusters they represent.
I1 and I5 further expressed desire to navigate to specific scales and clusters within a design.

I1: \textit{I'm really enjoying these links that I can kind of click on it...with the scales or clusters like they can take me to those. I was wondering...whether it would be possible to...maybe like pinpoint or just kind of go to the precise scale.}

I9: \textit{I was able to infer...there is one zoom level that has a particular region...and then they have a different zoom level that focuses on a different region and so on.}

Linking the analytics with design assemblages that they measure supported instructors in giving feedback to validate and refine what is measured. 
As instructors were able to inspect specific regions represented by analytics, they expressed where AI has a mismatch with their interpretation. 

I3: \textit{``I'm not sure why [it shows here] two different ones...you've got a couple [extra] clusters.''}

I1, I3, I4, and I9 derived assistance in understanding the analytics through our dashboard's visual annotation of the designs, which concretely represent the analytics through an animation of scales and clusters.
The animation\textemdash which presents clusters present at each scale one by one\textemdash helped instructors understand how design elements form spatial clusters across scales.
As I1 expressed, ``\textit{I think I now have a better understanding of spatial clusters [with] the animation of colors changing}''.

\subsection{Using Analytics for Assessment and Feedback}

The more design classroom sizes continue to grow, the more that instructors are challenged in having time to provide optimal levels of feedback to each student \cite{krause2017critique}.
Prior studies have found analytics useful toward scaling assessment and feedback \cite{pardo2019using,lim2019changes}.
Our findings for multiscale design analytics align.
I1 and I4 expressed that the analytics can become a part of their rubrics and feedback they give to students. 
Further, according to I4, making these analytics a part of rubrics can motivate and provide students guidance on what instructors are looking for in their design.

I1: \textit{I think [these analytics and my rubrics] complement each other. I think it will be very helpful...if there's a way that I can just sort of make a rubric on [dashboard] and attach to when they get their feedback.}

I4: \textit{You know, give them something to shoot for...I think that I would say...here are the things that I'd like to see in your design...I think that I would definitely like to assign scales as a part of the rubric to say, I would like to see the big picture from out here, and then when you zoom in, see more.}

We observe the potential of multiscale design analytics toward expediting instructors' assessment work.
I9 preferred to utilize analytics as a quick-to-use indicator of underlying problems. 
The analytics help them reduce the time they otherwise would spend on assessing each design. 

I9: \textit{So, I won't use the values in the column to directly give them points...But it's better than having to go to every [design] and look for every single issue or having a much larger rubric that I ran by...So think of the analytics as the symptoms and [then] you actually identify diseases.}

\subsection{Analytics as a Potential Source of Self-Reflection for Students}

Instructors (I1, I4, I5, I6) in our study expressed expectations that students would benefit from seeing multiscale design analytics.
According to them, seeing analytics has the potential to help students reflect on their progress.
More specifically, seeing analytics can help students in becoming aware of how they are organizing their ideas spatially across scales and clusters.
Self-reflection through analytics plays a vital role in learning, as it helps students in understanding their progress and stimulates improvements in their work \cite{wise2014designing,verbert2013learning}.

I5: \textit{I'm all for giving students as much information as they can use...and you know...they can use [analytics] to look at their progress.}

I1: \textit{Yeah, I would love students to explore more zoom levels...because usually, I think it is more like...I see it as an overall picture...but they don't really utilize being able to kind of go in to certain areas or zooming in to certain parts and elaborating...[Also,] maybe spatial clusters just so that they could be more aware about how they separate.}

While both I1 and I4 advocate for providing students with multiscale design analytics, they also caution against enforcing a specific type of visual organization. 
According to them, the goal of providing analytics would be to help students to reflect and effectively use multiscale organization, not to have a specific number of scales or clusters across scales.

I4: \textit{I wouldn't want them all to look the same like you don't want to go somewhere and see every painting looks the same, but it was almost as if some people were painting with boards and nails and hammers versus paintbrushes and paint. They just didn't really get what they're supposed to be putting on the [multiscale design]. So then it was just like not as effective.}

I1: \textit{[While] they have to become a little bit more mindful of [space]...just seeing how they lay out everything themselves...I would rather not control whether intentionally or unintentionally at all how they see spatial clusters.}

\section{Discussion + Implications: \\
Contextualizing Analytics to Support Design Education}
According to the model from Suchman's seminal treatise, \textit{Plans and Situated Actions} \cite{suchman1987plans}, the success of AI hinges on mutual intelligibility between AI and users; in our situated context, the users are instructors and students.
This mutual intelligibility depends on how the analytics, which function as linguistic expressions, get interpreted in the situated contexts of their use.
More specifically, multiscale design analytics are interpreted in the context of the situated instances of design work that students perform and the pedagogy and assessment that instructors provide.
We developed a research artifact to investigate how linking analytics to design instances would affect this interpretative process. 
We gathered and analyzed qualitative data to find out how instructors experience multiscale analytics when they are contextualized with this indexical linking to the design work they measure.

We first consider \textit{RQ2: What specific value can AI-based multiscale design analytics provide to design instructors in situated course contexts?}
We discuss and derive implications for how indexical presentation techniques, which link analytics to design instances, contextualize analytics and so support their use in abstract and creative tasks, here, in design education.

We then return to \textit{RQ1: How, if at all, can AI-based design analytics support instructors' assessment and feedback experiences in situated course contexts?}
Here, we discuss and derive implications regarding the particulars of multiscale analytics, what we've seen, and their potential to support assessment and feedback in design education.
We also consider limitations.

As Zimmerman et al. articulate, \textit{implications} are a form of theory produced using a Research through Design approach \cite{zimmerman2010analysis}; according to Gaver, the theory is likely to be ``provisional, contingent, and aspirational'' \cite{gaver2012should}.
Hence, we intend for investigating whether and how the implications from this study contribute to interfaces for deriving and presenting a range of complex analytics to be a fruitful avenue for future research. 
Such research can pinpoint, for example, whether particular implications are more useful in certain educational disciplines, in comparison to others.

\subsection{Indexicality: Demonstrating Design Analytics by Linking to Instances}
We contribute indexical linking from analytics to visually annotated design instances as a means of demonstrating what they mean.
According to Turnbull, indexical statements articulate relationships across contexts to convey new meanings  \cite{turnbull1993maps}.
In the present study, we found the indexicality of the dashboard\textemdash i.e., linking analytics with situated design element assemblages\textemdash supported instructors in understanding the analytics.
A key is the automatic visual annotation of a design to show which scales and clusters were recognized.
For example, in I9's words, \textit{``I was able to infer...there is one zoom level that has a particular region...and then they have a different zoom level that focuses on a different region and so on.''}
Instructors were able to get a quick sense of students' design organizations and how they were quantified.
They were also able to drill down to the work and pinpoint where the analytics mismatch their own interpretations.
For one of the designs, as I3 expressed on seeing AI results, \textit{``I'm not sure why [it shows here]...a couple [extra] clusters''}.

Other researchers are admirably pursuing explainable AI strategies for communicating the algorithmic logic of AI to users \citep[e.g.,][]{adadi2018peeking,samek2017explainable}.
This research alternatively contributes how linking analytics to instances can visually demonstrate to users what the analytics mean, showing the work of the recognition algorithm in situated contexts of practice, without attending to its underlying logic.

\textbf{\textit{Implications.} } 
Users, such as instructors and students, are expected to benefit when dashboard interaction directly indexes, that is, presents the linkages between specific analytics and design work instances. 
Suchman brings attention to how the transparency of AI-based systems\textemdash which is based in conveying an AI's intended purpose to users and establishing accountability\textemdash is vital for effectively supporting situated practice \cite{suchman1987plans}. 
A common AI approach is to get grades for a large set of assignments and build a bottom up recognizer from that data.
Such a recognizer is typically based on an arbitrary aggregation of features that can map to an overall grade score, rather than characteristics explicitly discernible to design instructors or students.
Alternatively, we crafted the multiscale design analytics of the present research, using contextual, design-based characteristics, to make sense to the design instructors. 
As a result, these analytics have the potential to inform students reflecting on how to improve their own work, as well as how to comprehend the work of others.

Further, our users found value in navigating to specific scales and clusters measured by analytics.
In I1's words, \textit{``whether it would be possible to...maybe like pinpoint or just kind of go to the precise scale.''}
For this, an intermediate representation on the dashboard can prove useful. 
For example, our findings motivate further investigation, in which beyond presenting one number on the dashboard, users are afforded interaction with a tree visualization \cite{estivill2002multi}, indexing the hierarchy of scales and clusters of each design work instance.
Such a representation has the potential to further support users in understanding how to use nested structures to convey complex information, going beyond the flatland to utilize a range of scales and clusters.

Lastly, in indexing analytics to instances, interfaces will benefit by using animations.
Mayer and Moreno showed that adding animations to study material enhances learner understanding \cite{mayer2002animation}. 
As Tversky explains, animations can aid perception and comprehension of the fine structure of spatial and temporal relationships among different pieces of content \cite{tversky2002animation}.
Bederson and Boltman found that animating viewpoint changes in a spatial environment helps users in building a mental map of the information present within the environment \cite{bederson1999does}. 
Multiscale design theory extends zoomable user interface theory to focus on how people assemble information elements in order to convey meanings, by using interaction based on space and scale and associated design principles.
In our study, we found that using animation helped instructors to understand complex characteristics.
In I1's words, ``\textit{I now have a better understanding of spatial clusters [with] the animation of colors changing}.''
We expect animation interactivity features, such as close-ups, zooming, and control of speed \cite{tversky2002animation} to prove useful in supporting navigation to specific scales and clusters.

\subsection{Supporting Assessment and Feedback in Design Courses through Multiscale Design Analytics}

For learning analytics to be effective in open-ended, project-based contexts, there is a need to assess complex characteristics that can give insights into students' creative strategies and abilities \cite{blikstein2011using}. 
Toward addressing this need, our study investigates how multiscale design analytics support instructors' assessment efforts in creative project-based learning contexts of design courses. 

Instructors in our study reported that multiscale design analytics can support them directly or indirectly in assessment and feedback processes. Instructors found that multiscale analytics have the potential to inform pedagogical intervention, based on whether or not students are able to effectively utilize the design environment.
In I9's words, \textit{``So if this number is extremely low for everybody...then maybe you need to [give] a tutorial [on the design environment].''}
Instructors shared that providing these analytics to students can help them reflect and improve their multiscale design skills.
For example, in I1's words, \textit{``I would love students to explore more zoom levels...they don't really utilize being able to...zooming in to certain parts and elaborating.''}

\textbf{\textit{Implications.} }
Our study demonstrates the potential of multiscale design analytics\textemdash which measure complex characteristics of design work\textemdash to assist instructors in assessing student work. 
The organization principles that design instructors expect their students to demonstrate map to the \textit{create} category in Bloom's revised taxonomy, i.e., ``put elements together'' into a ``new, coherent whole'' \cite{krathwohl2002revision,adams2015bloom}.
As I4 expressed, \textit{``I think that I would definitely like to assign scales as a part of the rubric to say, I would like to see the big picture from out here, and then when you zoom in, see more.''}
Likewise, I1 expressed that cluster analytics could help students reflect on their design representation and become \textit{``more aware about how they separate''}. 
We thus find that multiscale design analytics empower instructors in assessing students' holistic thinking and creative capabilities.

We advocate for future research that investigates further development of multiscale design analytics, visual annotations, and dashboard interactions, as well as more diverse and in-depth studies, in order to develop new knowledge about how to support instructors, in developing effective pedagogical interventions, and students, in learning how to do design that involves thinking about and presenting complex information.
This line of research has the potential to create new educational avenues for teaching how to present complex information that supports audiences in micro and macro readings\textemdash i.e., details and overviews \cite{cockburn2009review}\textemdash and the formation of mental models \cite{gentner2014mental} and maps.
Since conveying and understanding such information is vital in so many areas of society, this mission has the potential for broad impact that benefits society, through the work these students will perform throughout their careers.

Further, in design course contexts\textemdash where providing frequent feedback is vital\textemdash AI-based analytics demonstrate their utility in scaling the assessment.
For example, I9 finds using analytics \textit{``better than having to go to every [design] and look for every single issue or having a much larger rubric [to run] by.''}
Instructors in diverse project-based learning contexts\textemdash e.g., arts and humanities \cite{davies2011pbl}\textemdash engage students in creative, open-ended work.
Thus, these contexts are similarly expected to benefit from analytics based on assessments of complex characteristics.

The current research provides evidence for multiscale design measures to serve as \textit{descriptive analytics}, i.e., analytics that provide insights into student work \cite{uskov2019innovative}. Going ahead, with data from the past iterations of a course, these analytics have the potential to function as \textit{prescriptive analytics}, i.e., provide instructors and students alerts and suggestions based on computational modeling of the relationship between analytics and students' course performance \cite{arnold2012course,uskov2019innovative}.
On demand feedback through analytics has the potential to stimulate students' learning-by-doing.
Further, incorporating multiscale design analytics in widely distributed tools, such as Photoshop and Illustrator, has the potential to bring widespread benefits, as students use these tools in diverse design course contexts.

\subsection{Limitations of Multiscale Design Analytics}

Multiscale design analytics are not a panacea.
On the one hand, our findings show that the present multiscale design analytics provide value to instructors, in situated course contexts.
I9 would like to see that students are able to effectively use the multiscale design environment.
I2 values students' presentation of structure at different levels.
I1 would \textit{``love students to explore more zoom levels''}, as she does not see them \textit{``zooming in to certain parts and elaborating''}.

On the other hand, multiscale design analytics were not found to serve as a catch-all measure for design.
I4 talks about not wanting all designs to look the same \textit{``like you don't want to go somewhere and see every painting looks the same''}. 
I1 says, \textit{``I would rather not control...how they see spatial clusters''}.

We build theory using creative cognition's family resemblance principle, according to which, no particular characteristics are required for a work to be deemed creative \cite{smith1995creative}. 
Rather, a family of traits tends to serve as indicative.
We find that multiscale design, as measured here, functions as one such design creativity trait.
As another fruitful avenue for future research, we identify deriving analytics for families of design creativity traits\textemdash for example, feasibility, originality, and aesthetics \cite{adams2016characterizing,christensen2016dimensions}; and gestalts principles, e.g., proximity, closure, continuity, symmetry, parallelism, and similarity of color, size, and orientation \cite{wagemans2012century}\textemdash and applying these traits in education and even crowd-sourced design contexts.
According to the family resemblance principle, as no particular trait is sufficient, design creativity analytics will never be perfect. 
But inasmuch as they work well enough, they can provide instructors, students, and other designers with insights so as to (1) provide first-order assessment; and (2) stimulate ongoing work.

\section{Conclusion}
We took a Research through Design approach and created a research artifact to understand the implications of AI-based multiscale design analytics, in practice.
Our study demonstrates the potential of multiscale design analytics for providing instructors insights into student design work and so support their assessment efforts.
We focused on supporting users engaged in creative design tasks.
Underlying our investigation was our understanding of how multiscale design contributes to teaching and performing these tasks.

We develop multiscale design theory to focus on how people assemble information elements in order to convey meanings.
The tasks that students perform in the assignments cross fields.
Multiscale design tasks are exploratory search tasks, which involve looking up, learning, and investigating \cite{marchionini2006exploratory}.
They are information-based ideation tasks, which involve finding and curating information elements in order to generate and develop new ideas as part of creativity and innovation \cite{kerne2014using,jain2015tweetbubble}.
They are visual design thinking tasks, which involve forming combinations through sketching and the reverse, sketching to generate images of forms in the mind \cite{goldschmidt1994visual}.
They are constructivist learning tasks, in which making serves as a fundamental basis for learning by doing \cite{jonassen1994thinking,yager1991constructivist,blumenfeld1991motivating}.
On the whole, multiscale design has roots in diverse fields and, as we see from our initial study, applications in diverse fields.
The scopes of intellectual merit and potential broad impact are wide.

The present research contributes how to convey the meaning of multiscale design analytics derived using AI, by linking dashboard presentation of design analytics with the actual design work that they measure and characterize.
Making AI results understandable by humans is fundamental to building their trust in using systems supported by AI \cite{shneiderman2020human}.
In our study, when the interface presents what is being measured by AI, it allows users to agree or disagree.
Specifically, our integration of the dashboard presentation with the actual design environment allowed instructors to independently validate the particular sets of design element assemblages that the AI determined as nested clusters.
This makes the interface to the AI-based analytics visible, or as Bellotti and Edwards said, intelligible and accountable \cite{bellotti2001intelligibility}.
The importance of making AI decisions visible has been noted in healthcare \cite{wang2019designing,bussone2015role} and criminal justice \cite{deeks2019judicial} domains.
Likewise, in education, supporting users' understanding of AI-based analytics is vital, as the measures can directly impact outcomes for an individual.
Analytics that do not connect with students' design work would have little meaning for instructors, if at all.
Students, if provided with such analytics, would fail to understand and address the shortcomings that they indicate.

Significant implications for future research are stimulated by the current level of investigation of the particular multiscale design analytics in particular situated course context classrooms.
We need further investigation of how these as well as new multiscale design analytics affect other design education contexts and design in industry.
Such research can investigate the extent to which different analytics and visualization techniques\textemdash e.g., indexical representation and animation\textemdash are beneficial in specific contexts.
Actionable insights on design work can prove vital in improving learners’ creative strategies and abilities, which in turn can stimulate economic growth and innovation \cite{NAE2010}.
Continued efforts toward simultaneously satisfying the dual goals of AI performance and visibility of decisions\textemdash across a range of contexts\textemdash has the potential to create broad impacts by providing inroads to addressing complex sociotechnical challenges, such as ensuring reliability and trust \cite{shneiderman2020human} in the use of AI systems.

\bibliographystyle{ACM-Reference-Format}
\bibliography{references}

\appendix

\section{Interview Questions}
\label{sec:questions}
We used the following questions to guide our semi-structured interviews:

\begin{itemize}[leftmargin=*]
\small
\item Please briefly describe your experiences with the courses dashboard.

\item Do you think the class would be different with and without the dashboard? If so, how?

\item How does how you use the courses dashboard compare with other learning management systems and environments? What is similar? Is anything different?

\item Has using the dashboard shown you anything new or unexpected about your students' learning? If yes, what?

\item What do you understand about the analytics presented on the dashboard with submissions?

\item Do you utilize analytics? If so, do they support in monitoring and intervening? Assessment and feedback? How? 
\item If the answer to \textit{`Do you utilize analytics'} is \textit{`No'}: Do you think these analytics have the potential to become a part of the assessment and feedback that you provide to the students? If so, how?

\item What do you think about showing these analytics to students on-demand?

\item Did you click on `Scales' analytics? How did seeing its relationship with the actual design work affect your utilization (or potential utilization) for assessment and feedback?

\item Did you click on `Clusters' analytics? How did seeing its relationship with the actual design work affect your utilization (or potential utilization) for assessment and feedback?

\item Has using the dashboard to follow and track student design work changed how you teach or interact with the students? If so, how?

\item What would you do different, if anything, next time you teach the class?

\item What are your suggestions for making the dashboard more suited for your teaching and assessment practices? Or for design education in general?

\end{itemize}

\end{document}